\documentclass{aa}  

\usepackage{graphicx}
\usepackage{txfonts}
\usepackage{color}
\usepackage[normalem]{ulem}

\begin{document}

   \title{Quantifying the fine  structures of disk galaxies with deep learning:
Segmentation of spiral arms
in different Hubble types}
\titlerunning{Segmentation of spiral arms}


   \author{K. Bekki 
          \inst{1}
          }

   \institute{ ICRAR,
M468,
The University of Western Australia
35 Stirling Highway, Crawley
Western Australia, 6009, Australia \\
              \email{kenji.bekki@uwa.edu.au}
             }

   \date{Received September 15, 1996; accepted March 16, 1997}

 
  \abstract
   {
Spatial correlations between spiral arms and other galactic components
such as giant molecular clouds and massive OB stars
suggest that
spiral arms can play vital roles in various aspects of disk galaxy
evolution. Segmentation of spiral arms in disk galaxies is therefore a key task when these correlations are to be investigated.
   }
   {
We therefore decomposed disk galaxies into spiral and nonspiral regions
using the code U-Net, which is based on deep-learning algorithms
and has been invented for segmentation tasks in biology.
  }
{
 We first trained this U-Net with a large number  of synthesized images
of disk galaxies with known  properties of symmetric spiral arms
with radially constant pitch angles
and then tested it with entirely unknown data sets.
The synthesized images were generated from mathematical models of
disk galaxies with various properties of spiral arms, bars, and rings
in these supervised-learning tasks.
We also applied the trained U-Net to spiral galaxy images synthesized
from the results of long-term hydrodynamical simulations of disk galaxies
with nonsymmetric spiral arms.
}
{
We find that U-Net can  predict the precise locations of spiral arms
with an average prediction accuracy ($F_{\rm m}$) of 98\%.
We also find that
$F_{\rm m}$  does not depend strongly on the numbers of spiral arms,
presence or absence of stellar bars and rings,
and bulge-to-disk ratios in disk galaxies.
These results imply that U-Net is a very useful tool for identifying
the locations of spirals arms. However, we find 
that the U-Net trained on these symmetric spiral arm images cannot predict  entirly unknown data sets with the same accuracy that were produced from
the results of hydrodynamical simulations of disk galaxies with
nonsymmetric irregular spirals and their nonconstant  pitch angles
across disks. In particular, weak spiral arms 
in barred-disk galaxies
are properly segmented.
}
   {
These results suggest that U-Net can segment 
more symmetric spiral arms with constant pitch angles in disk galaxies.
However, we need 
to train U-Net with a larger number of more realistic galaxy images
with noise, nonsymmetric spirals,
and different pitch angles between different arms in order to apply
it to real spiral galaxies.
It would be a challenge to make a large number of training data sets  
for such realistic nonsymmetric and irregular spiral arms with nonconstant
pitch angles.
}


   \maketitle
%

\section{Introduction}

Spiral arms play various roles in disk galaxy evolution through their
gravitational effects on disk field stars and gaseous dissipation
associated with their hydrodynamical interaction  with the interstellar medium (ISM).
For example, cold gaseous components of disk galaxies can
be strongly perturbed by gravitational potentials of spiral
arms so that their densities and motions can be significantly changed
(e.g., Fujimoto 1968; Roberts 1969; Egusa et al. 2017).
Formation processes  of giant molecular clouds (GMCs) 
within spiral arms  and
the mass functions of GMCs depend on the spiral arm properties.
(e.g., Tasker et al. 2015; Pettitt et al. 2020).
Resonant dynamical interaction between spiral arms and disk field stars
at corotation radii in disk galaxies can result in an
angular momentum redistribution and the resultant radial migration of stars
(e.g., Sellwood \& Binney 2002).
Dynamical heating of stars by spiral arms themselves can increase
the velocity dispersion of stars and thus cause the weakening of spiral arms (Sellwood \& Carlberg 1984; but see
Fujii et al. 2011; D'Onghia et al. 2013 for
more recent results on the longevity of spiral arms).

Many observational studies quantified the detailed properties of spiral arms,
for instance, numbers,  pitch angles, shapes (e.g., grand design or flocculent),
and amplitudes (e.g., Elmegreen \& Elmegreen 1984; Rix \& Zaritsky 1995,
Seigar \& James 1998; Kendall et al. 2008; Davis et al. 2012)
and also revealed physical correlations between these arm properties
and other galactic properties such as bulge-to-disk ratios
and massive black hole masses
(e.g., Kennicutt 1981; Seiger et al. 2008; Davis et al. 2017).
Previous theoretical studies and numerical simulations indeed demonstrated that
these physical properties of spiral arms 
can depend on the physical properties
(e.g., Q parameters) of galactic disks because
the effectiveness of spiral formation mechanisms are controlled by the 
disk properties
(e.g., Athanassoula 1984; Carlberg \& Freedman 1985; Grand et al. 2013;
see Dobbs \& Baba 2014 for a recent review). 
Recently, statistical studies of spiral arm properties based on a large
number of galaxy images from the Galaxy Zoo project
have been made, which provides new clues for 
the origin of spiral arms in disk galaxies (e.g., Masters et al. 2019).
These observed properties of spiral arms have not only constrained
competing theoretical models of spiral arms 
(e.g., Sellwood 2011), but also assisted astronomers in
understanding various roles of spiral arms in galaxy evolution
(e.g., Seiger et al. 2008).
Thus, the details of  spiral properties quantified for a large
number of disk galaxies with different Hubble morphological types 
will enable us to make even more significant progress in understanding the 
formation processes of spirals and their physical roles in galaxy evolution.

   \begin{table}
      \caption[]{Description of physical meanings for symbols.}
         \label{KapSou}
     $$ 
         \begin{array}{ll}
            \hline
            \noalign{\smallskip}
{\rm parameters}  & {\rm  meanings} \\
            \noalign{\smallskip}
            \hline
            \noalign{\smallskip}
N & {\rm Number  \hspace{1mm}  of  \hspace{1mm}  spiral \hspace{1mm}  arms } \\
f_{\rm sp} & {\rm Mass \hspace{1mm}   fraction \hspace{1mm}  of \hspace{1mm}  spiral \hspace{1mm}  arms} \\
w_{\rm sp} & {\rm Spiral  \hspace{1mm} arm \hspace{1mm}  width} \\
\theta_{\rm max} &  {\rm Maximum \theta \hspace{1mm}  in  \hspace{1mm}  logarithmic \hspace{1mm}  spiral \hspace{1mm}  models} \\
f_{\rm bul} & {\rm Bulge \hspace{1mm} mass \hspace{1mm} fraction} \\
f_{\rm bar} & {\rm Bar \hspace{1mm} mass \hspace{1mm} fraction} \\
R_{\rm bar} & {\rm Bar\hspace{1mm}  size} \\
A_{\rm bar} & {\rm Bar \hspace{1mm} axis \hspace{1mm} ratio} \\
F_{\rm m} & {\rm Prediction \hspace{1mm} accuracy \hspace{1mm} of \hspace{1mm}  N-Net} \\
F_{\rm m, sp} & F_{\rm m} {\rm for \hspace{1mm}  spiral \hspace{1mm} arm \hspace{1mm}  regions} \\
F_{\rm m, nsp} & F_{\rm m} {\rm for \hspace{1mm}  non-spiral \hspace{1mm}  arm \hspace{1mm}  regions} \\
\Sigma  & {\rm Projected \hspace{1mm}  stellar \hspace{1mm}  mass \hspace{1mm}  density} \\
\theta & {\rm Disk\hspace{1mm}  inclination \hspace{1mm}  angle \hspace{1mm}  w.r.t. \hspace{1mm}  disk\hspace{1mm}  spin \hspace{1mm}  axis}\\
f_{\rm ring} & {\rm Ring \hspace{1mm}  mass \hspace{1mm}  fraction} \\
R_{\rm ring} & {\rm Ring\hspace{1mm}  size \hspace{1mm}  (=R_{\rm bar}) } \\
R_{\rm d} & {\rm Disk \hspace{1mm}  size}  \\
R_{\rm f} & {\rm Image \hspace{1mm}  size \hspace{1mm}  w.r.t. \hspace{1mm}  R_{\rm d} } \\
            \noalign{\smallskip}
            \hline
         \end{array}
     $$ 
\end{table}

Future large surveys of galaxies based on next-generation observational facilities
such as LSST, TMT, and EUCLID will generate a huge number of galaxy images in the nearby
and distance universe: for example,
the LSST will be able to observe about 20 billion  galaxies.
Because spiral arms and bars are the main galactic structures
that drive galaxy evolution (e.g., Buta 2013),
these future observations will provide an excellent large data set for
the various properties of spiral galaxies. 
Automated classification and quantification
of spiral arms in disk galaxies will therefore greatly assist astronomers in understanding
the physics of the dynamical evolution of spiral galaxies  based 
on the statistical properties of the galaxies.
Davis \& Hayes (2014) have recently developed a new automated spiral extraction
method and applied it to nearly 30,000 galaxies taken from the
Galaxy Zoo project (Lintott et al. 2008). Their methods for spiral arm
segmentation are not based on
convolutional neural networks (CNNs), which  have been widely used in recent 
image analyses of galaxies (e.g.,Dieleman et al. 2015; Huertas-Company et al. 2015;
 Dominguez Sanchez et al. 2018).

   \begin{figure}
   \centering
   \includegraphics[width=8.5cm]{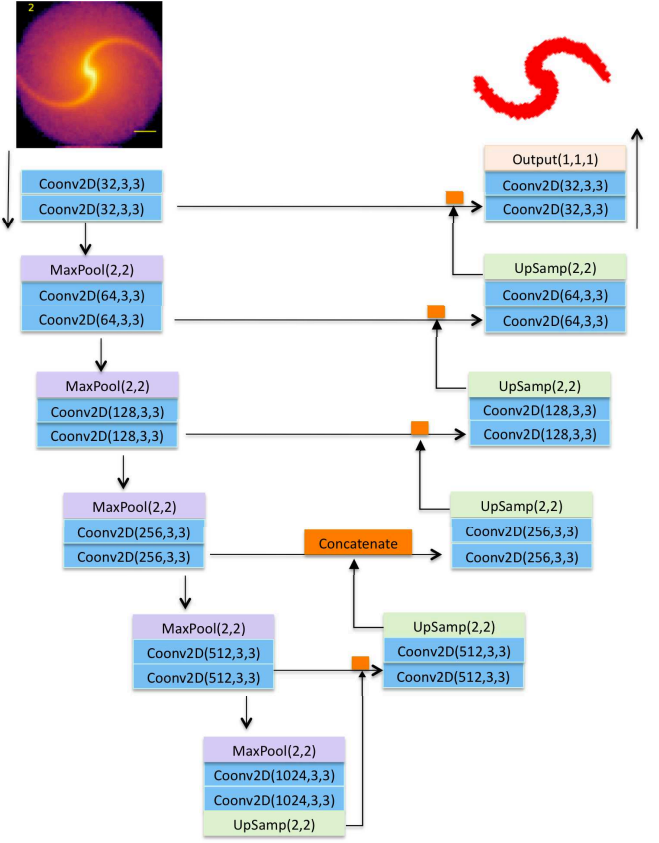}
      \caption{
U-Net architecture adopted in the present study. Rectangular
boxes with different colors
represent 2D convolutional (blue), maximum pooling (purple),
up-sampling (green), and output (orange) layers.
The input image (2D density map of a spiral galaxy)
is processed in these layers counterclockwise along this U-shaped
architecture, and
the dimensions of input and output data  (channel numbers, etc.)
are shown within  each box for each layer.
The two $3 \times 3$ 2D
convolutional layers with a ReLU each
are followed by a $2 \times 2$ maximum pooling layer,
and this image processing is repeated at the left (contracting).
In the right (expanding) part of U-Net,
the output from a convolutional layer is concatenated with
that from an up-sampling layer, and this concatenation
is indicated by orange rectangles.
For example,
two 2D convolutional layers with 128 channels within the third block
from top in the right part of U-Net are followed by an up-sampling layer,
and
the output of the up-sampling layer is concatenated with
the output from the second convolutional layer with 64 channels
within the second block from top
in the left part of U-Net,
and then input into the first convolutional layer with 64 channels within
the second block from the top at the right.
A sigmoid activation function rather than softmax is used in the final
output layer in the present segmentation tasks of spiral arms.
              }
         \label{FigVibStab}
   \end{figure}

   \begin{table*}
      \caption[]{Summary 
of the parameter values in training data sets for U-Net.}
         \label{KapSou}
     $$ 
         \begin{array}{llllllllll}
            \hline
            \noalign{\smallskip}
{\rm U-Net \hspace{1mm}  ID} & N & f_{\rm sp}  & w_{\rm sp} & \theta_{\rm max} & f_{\rm bul} &
f_{\rm bar} & R_{\rm bar} & F_{\rm m}
& {\rm comments } \\
            \noalign{\smallskip}
            \hline
            \noalign{\smallskip}
U1 & 2 & 0.4-0.9  & 0.03 & 3.14-6.28 & 0-0.1 & 0 & - & 0.987 & {\rm fiducial \hspace{1mm}  model} \\
U2 & 1-5 & 0.4-0.9  & 0.03 & 3.14-6.28 & 0-0.1 & 0 & - & 0.982 & {\rm  multiple\hspace{1mm}  arms} \\
U3 & 1-5 & 0.4-0.5  & 0.03 & 3.14-6.28 & 0.1-0.5 & 0 & - & 0.965 & {\rm larger \hspace{1mm}  bulges} \\
U4 & 1-5 & 0.1-0.5  & 0.03 & 3.14-6.28 & 0.1-0.5 & 0.4 & 0.5 & 0.985 & {\rm bars\hspace{1mm}  and
\hspace{1mm}  larger \hspace{1mm}  bulges} \\
U5 & 1-5 & 0.1-0.5  & 0.03-0.06 & 3.14-6.28 & 0-0.1 & 0.4 & 0.5 & 0.952 & {\rm bars \hspace{1mm}   and \hspace{1mm}   wider \hspace{1mm}  spirals} \\
U6 & 1-8 & 0.4-0.9  & 0.03 & 3.14-6.28 & 0-0.1 & 0 & - & 0.977 & {\rm large \hspace{1mm}  arm  \hspace{1mm}  number} \\
            \noalign{\smallskip}
            \hline
         \end{array}
     $$ 
\end{table*}
A new architecture of CNNs (U-Net)
for biomedical image-segmentation tasks has
recently been developed (Ronneberger et al. 2015)
and is now widely used for various purposes of image analysis
 not only in biology, but also in other areas of research.
In spite of its power and great performance in image segmentation tasks,
U-Net has not been widely used in galaxy image analysis so far.
Boucaud et al. (2019) applied U-Net to galaxy images from the CANDELS survey
in order to separate two (or multiple) galaxies from blended galaxy systems (e.g., 
interacting galaxies or similar apparent locations of two galaxies at different z
by chance). If U-Net can separate spiral arms from
other axisymmetric (e.g., exponential disks)  or nonaxisymmetric (e.g., bars)
components in disk galaxies, then astronomers can investigate
whether the locations of spiral arms are correlated with intense star-forming
regions or the locations of giant molecular clouds (GMCs) and thereby
gain better understanding of galaxy-scale star formation and 
GMC formation processes triggered by
spiral arms.

The purpose of this paper is to locate spiral arms in disk galaxies with
different Hubble morphological types using U-Net.
In this first paper, we assess the capability of U-Net
to distinguish between spiral and nonspiral regions of disk galaxies  
using a larger number of synthesized images from initial 
spiral galaxies  in computer simulations (i.e., not
observed images of spirals).
These clean
images do not contain any noise, nearby stars, or external disturbances
from galaxy interaction and environmental effects so that we can better assess
the capability of U-Net for spiral segmentation task.
If U-Net can properly identify and locate spiral arm structures 
in these clean images,
then we can conclude
that 
U-Net  is a promising tool for spiral segmentation
and thus will be able to be used  for spiral segmentation
in real images of galaxies or  synthesized images of galaxies from more complicated
cosmological simulations of galaxy formation.
If it cannot accomplish this, we would need to look
for alternative ways for spiral arm segmentation.
Our previous studies used CNN to find fine structures
in S0s (Diaz et al. 2019), barred spiral galaxies (Cavanagh \& Bekki 2020),
and outer gas disks of spiral galaxies (Shah et al. 2019). The CNN
architectures (U-Net) used in the present work are quite different from those
used in these previous works.


   \begin{table*}
      \caption[]{Summary of parameter values in testing data sets for U-Net.
}
         \label{KapSou}
     $$ 
         \begin{array}{lllllllllllll}
            \hline
            \noalign{\smallskip}
{\rm model \hspace{1mm}  ID} & N & f_{\rm sp}  & w_{\rm sp} & \theta_{\rm max} & f_{\rm bul} &
f_{\rm bar} & R_{\rm bar} & A_{\rm bar}  & R_{\rm ring} & f_{\rm ring} &
R_{\rm f}  & F_{\rm m} \\
            \noalign{\smallskip}
            \hline
            \noalign{\smallskip}

T1 & 2 & 0.1-0.5  & 0.03 & 3.14-6.28 & 0.1-0.5 & 0 & - & - & - & - & 1.0 & 0.962 \\
T2 & 2 & 0.4-0.5  & 0.03 & 3.14-6.28 & 0.1-0.5 & 0 & - & - & - & - & 1.0 & 0.966\\
T3 & 2 & 0.1-0.4  & 0.03 & 3.14-6.28 & 0-0.1 & 0 & - & - & -& -& 1.0 & 0.940\\
T4 & 2 & 0-0.05  & 0.03 & 3.14-6.28 & 0-0.1 & 0 & - & - & -& -& 1.0 & 0.761\\
T5 & 2 & 0.05-0.1  & 0.03 & 3.14-6.28 & 0-0.1 & 0 & - & - & -& -& 1.0 & 0.809\\
T6 & 2 & 0.1-0.2  & 0.03 & 3.14-6.28 & 0-0.1 & 0 & - & - & -& -& 1.0 & 0.887 \\
T7 & 2 & 0.2-0.3  & 0.03 & 3.14-6.28 & 0-0.1 & 0 & - & - & -& -& 1.0 & 0.954\\
T8 & 2 & 0.4-0.9  & 0.03 & 6.29-9.42 & 0-0.1 & 0 & - & - & - & -& 1.0 & 0.979\\
T9 & 5 & 0.4-0.9  & 0.03 & 6.29-9.42 & 0-0.1 & 0 & - & - & - & -& 1.0 & 0.936\\
T10 & 2 & 0.4-0.5  & 0.03 & 3.14 & 0-0.1 & 0.4 & 0.5 & 0.2 & - & -& 1.0 & 0.958\\
T11 & 2 & 0.4-0.5  & 0.03 & 3.14 & 0-0.1 & 0.4 & 0.5 & 0.6 & - & -& 1.0 & 0.959\\
T12 & 2 & 0.4-0.5  & 0.03 & 3.14 & 0-0.1 & 0.4 & 0.2 & 0.2 & - & - &1.0 & 0.983 \\
T13 & 2 & 0.4-0.9  & 0.06 & 3.14-6.28 & 0-0.1 & 0 & - & - & - & - &1.0 & 0.898\\
T14 & 2 & 0.4-0.9  & 0.02 & 3.14-6.28 & 0-0.1 & 0 & - & - & - & - &1.0 & 0.974\\
T15 & 6 & 0.4-0.9  & 0.03 & 3.14-6.28 & 0-0.1 & 0 & - & - & - & - & 1.0 & 0.934 \\
T16 & 7 & 0.4-0.9  & 0.03 & 3.14-6.28 & 0-0.1 & 0 & - & - & - & - & 1.0 & 0.973\\
T17 & 2 & 0.4-0.5  & 0.03 & 3.14 & 0-0.1 & 0.3 & 0.5 & 0.2 & 0.5 & 0.1 & 1.0 & 0.903\\
T18 & 4 & 0.4-0.5  & 0.03 & 3.14 & 0-0.1 & 0.3 & 0.5 & 0.2 & 0.5 & 0.1 & 1.0 & 0.919\\
T19 & 2 & 0.4-0.5  & 0.03 & 3.14 & 0-0.1 & 0.3 & 0.2 & 0.2 & 0.2 & 0.1 & 1.0 & 0.976\\
T20 & 4 & 0.4-0.5  & 0.03 & 3.14 & 0-0.1 & 0.3 & 0.2 & 0.2 & 0.2 & 0.1 & 1.0 & 0.972\\
T21 & 2 & 0.4-0.9  & 0.03 & 3.14-6.28 & 0-0.1 & 0 & - & - & - & - &1.2 & 0.979\\
T22 & 2 & 0.4-0.9  & 0.03 & 3.14-6.28 & 0-0.1 & 0 & - & - & -& - & 1.5 & 0.971 \\
T23 & 2 & 0.4-0.9  & 0.03 & 3.14-6.28 & 0-0.1 & 0 & - & - & -& - &2.0 & 0.952\\
            \noalign{\smallskip}
            \hline
         \end{array}
     $$ 
\end{table*}

\section{Model}

The adopted  mass
distributions of stellar disks and bulges are the same as we used
in our previous N-body and hydrodynamical simulations of disk galaxies
(e.g., Bekki 2014a, 2015).
We generated a large number of disk galaxy models with different properties
of spiral arms represented by numerous particles, as done in our
previous simulations.
The mass distributions of spiral arms are based on logarithmic spiral arm models,
which are more mathematical.
These  models of spiral galaxies can generate a large ($\sim 50000$ in total)
number of clean images of galaxies, which we think suitable for
assessing the capability of U-Net of segmenting spiral arms in
disk galaxies. 

\subsection{Stellar disks and bulges}

A spiral galaxy is assumed to consists of dark matter halo, stellar disk,
and spherical stellar  bulge (no gas)
in the present disk galaxy models.
Because dark matter halos and kinematics of disk galaxies
are not investigated at all in this study,
we model only  the structures of spiral galaxies.
The total masses of stellar disk and
and stellar bulge in a disk galaxy 
are denoted as $M_{\rm d}$
and $M_{\rm bul}$, respectively, and
the total stellar mass of a disk galaxy is denoted by $M_{\rm gal}$.
The spherical stellar bulge of a disk galaxy  has a size of $R_{\rm bul}$
and a scale-length of $R_{\rm 0, bul}$ ($=0.2R_{\rm bul}$)
and is represented by the Hernquist
density profile. The stellar disk
with a size of $R_{\rm d}$  is represented by 
an exponential profile and 
the radial ($R$) and vertical ($Z$) density profiles  are
therefore proportional to $\exp (-R/R_{0}) $ with scale
length $R_{0} = 0.2R_{\rm d}$  and to ${\rm sech}^2 (Z/Z_{0})$ with scale
length $Z_{0} = 0.04R_{\rm d}$, respectively.

\subsection{Logarithmic spirals,  bars, and rings}

The stellar disk is assumed to have spiral arms that are represented by
logarithmic spiral arms  adopted in previous models (e.g., Danver 1942;
Ma 2001;  Seigar \& James 1998;
Davis et al. 2012, Garcia-Gomez et al. 2017),
and a spiral arm  is described
in the polar coordinate $(r, \theta)$ as follows:
\begin{equation}
r=a_0e^{b_0 \theta},
\end{equation}
where $a_0$ and $b_0$ are the parameters that control the 
the shape of a spiral arm. Because we consider that the spiral arms in a disk galaxy
all extend to $R=R_{\rm d}$ (disk size), the maximum $\theta$ ($\theta_{\rm max}$) should
satisfy the following relation:
\begin{equation}
R_{\rm d}=a_0e^{b_0 \theta_{\rm max}},
\end{equation}
which gives the relation of $a_0$ and $b_0$. Some models have stellar bars, so that
the above equation needs to be revised as follows:
\begin{equation}
R_{\rm d}-R_{\rm bar}=a_0e^{b_0 \theta_{\rm max}},
\end{equation}
where $R_{\rm bar}$ is the size of a stellar bar. We mainly investigate
the models with $\theta_{\rm max}$ ranging from 3.14 and 9.42 (i.e., $\pi$ to $3\pi$)
for $b_0=0.5$.
The mass density of a spiral arm at $R$ follows the adopted exponential density
profile at $R$. The width of a spiral arm is represented by $w_{\rm sp}$, and it
is assumed to 
range from 0.02 to 0.06$R_{\rm d}$.
The mass fraction of spiral arms in a disk galaxy ($f_{\rm sp}=M_{\rm sp}/M_{\rm gal}$,
where $M_{\rm sp}$ is the total mass of the arms) is a free parameter that defines
the strength (or the amplitude) of the arms. Each  of multiple spiral arms is assumed
to have  the same shapes and mass distributions
 in this study, although this assumption is not realistic.

A stellar bar is modeled as a Ferrer two-dimensional bar with $n=1,$ and the boundary
$d (x, y)$ within which stars are considered to belong to the bar is described as 
follows:
\begin{equation}
d(x,y)=\left(\frac{x}{ a_{\rm bar} } \right)^2+\left (\frac{y}{ b_{\rm bar} }
\right)^2,
\end{equation}
where $a_{\rm bar}$ and $b_{\rm bar}$ are the lengths of the major and minor axes,
respectively. If this $d(x,y)$ is less than 1 at a particle position
$(x,y)$, then the particle is regarded to lie within a bar.
It should be stressed that this equation defines the boundary of the stellar bar,
and a particular type
of the particle distribution within the bar (e.g., exponential profile)
can still be chosen.
The mass density distribution within a bar of a disk galaxy is here assumed to follow the adopted
exponential profile of the disk.
The axis ratio of a bar ($A_{\rm bar}=b_{\rm bar}/a_{\rm bar}$) is assumed to be
a key free parameter that controls the shape of the bar.
This model is the same as was adopted by Garcia-Gomez et al. (2017),
although we investigate the Ferrer models with $n=1$ only.
The mass fraction of a stellar bar
 in a disk galaxy ($f_{\rm bar}=M_{\rm bar}/M_{\rm gal}$,
where $M_{\rm bar}$ is the total mass of the bar)
is a free parameter that defines
the strength of the bar.

We also constructed the barred spiral galaxy models  with inner rings around
bars in order to investigate whether U-Net can distinguish
between spiral arms with low pitch angles and inner rings.
This is a tough test, but we can better understand the capability
of U-Net of segmenting spiral arms in disk galaxies using these galaxy images
with bar, spirals, and rings. We assumed that the size of a ring
($R_{\rm ring}$)  in
a barred spiral galaxy is the same as the bar length ($R_{\rm bar}$).
The ring was assumed to be circular with a width of $0.05R_{\rm d}$ and a uniform
mass distribution within the ring.
The mass fraction of a ring ($M_{\rm ring}/M_{\rm gal}$, where
$M_{\rm ring}$ is the total mass of the ring)
was assumed to be fixed at 0.1 and  is represented by $f_{\rm ring}$.
We investigated  barred spiral galaxies with or without rings around 
bars. The brief descriptions of these model parameters for disk galaxies 
are given in Table 1.

\subsection{U-Net}

U-Net was developed for precise and efficient segmentation of biomedical
images (Ronneberger et al. 2015) and has been widely used for various purposes.
We here use U-Net to segment spiral arms of disk galaxies as precisely as
possible. We applied U-Net to spiral galaxy images
after we modified
 the final layer of the original  architecture (Ronneberger et al. 2015).
The details of the adopted architecture with layer types, input dimensions, weight shapes,
activation functions, etc. are described in Fig. 1.
We show the results only from  this architecture in Fig. 1 because it can 
precisely segment the spiral arms in disk galaxies, as we show below. We do not
intend to discuss how the different architectures can improve the 
precision of spiral segmentation in this study.

Because we adopted the sigmoid  (rather than softmax) activation function
(which outputs real numbers ranging from 0 to 1, i.e., not 0 or 1 for
classification) in the  final layer,
we considered how each pixel might be classified into a spiral or nonspiral region.
We considered that the output value for each pixel is higher than
a threshold value ($P_{\rm th}$). The pixel can then be classified
as a spiral region. We mainly investigated models with $P_{\rm th}=0.5$
because the results do not depend strongly on $P_{\rm th}$.
We trained U-Net for 30 epochs for each set of models because 30 epochs are enough
for the adopted U-Net to be well trained for a precise segmentation ($F_{\rm m}>0.97$).
Table 2 summarizes the six sets of disk galaxies models that we used to generate
the synthesized images of galaxies on which we trained U-Net.
In order to train and test U-Net, 
we used the publicly available code Keras (Chollet 2015) with a collection
of neural network libraries for deep learning.
We trained U-Net with the NVIDIA GPU GTX1080 on the Magellan
GPU cluster at the University of Western Australia. 

   \begin{figure*}
   \centering
  \includegraphics[width=18cm]{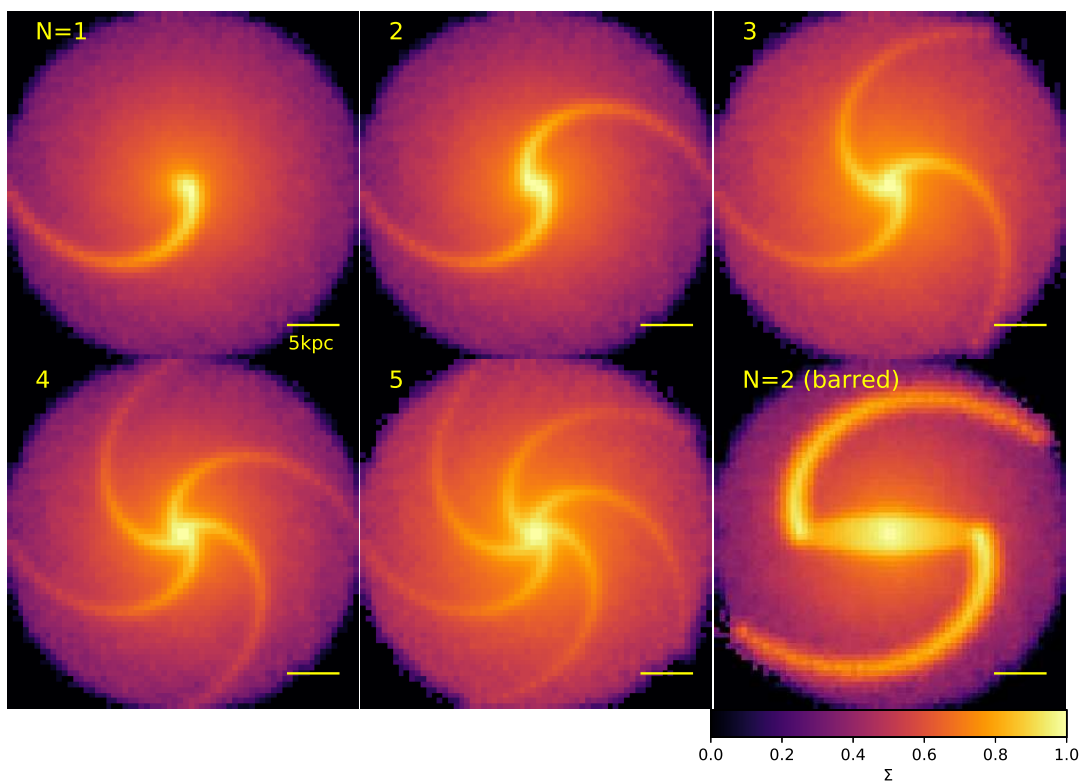}
   \caption{
Two-dimensional maps of surface mass densities ($\Sigma$ on logarithmic scale)
of spiral galaxies
projected onto the $x$-$y$ plane for six different models with different
$N$ and with and without stellar bars. The 2D densities
are normalized ($0 \le \Sigma \le 1$).The disk size in these galaxy images
is assumed to be 17.5 kpc.
               }
              \label{FigGam}%
    \end{figure*}

   \begin{figure*}
   \centering
  \includegraphics[width=18cm]{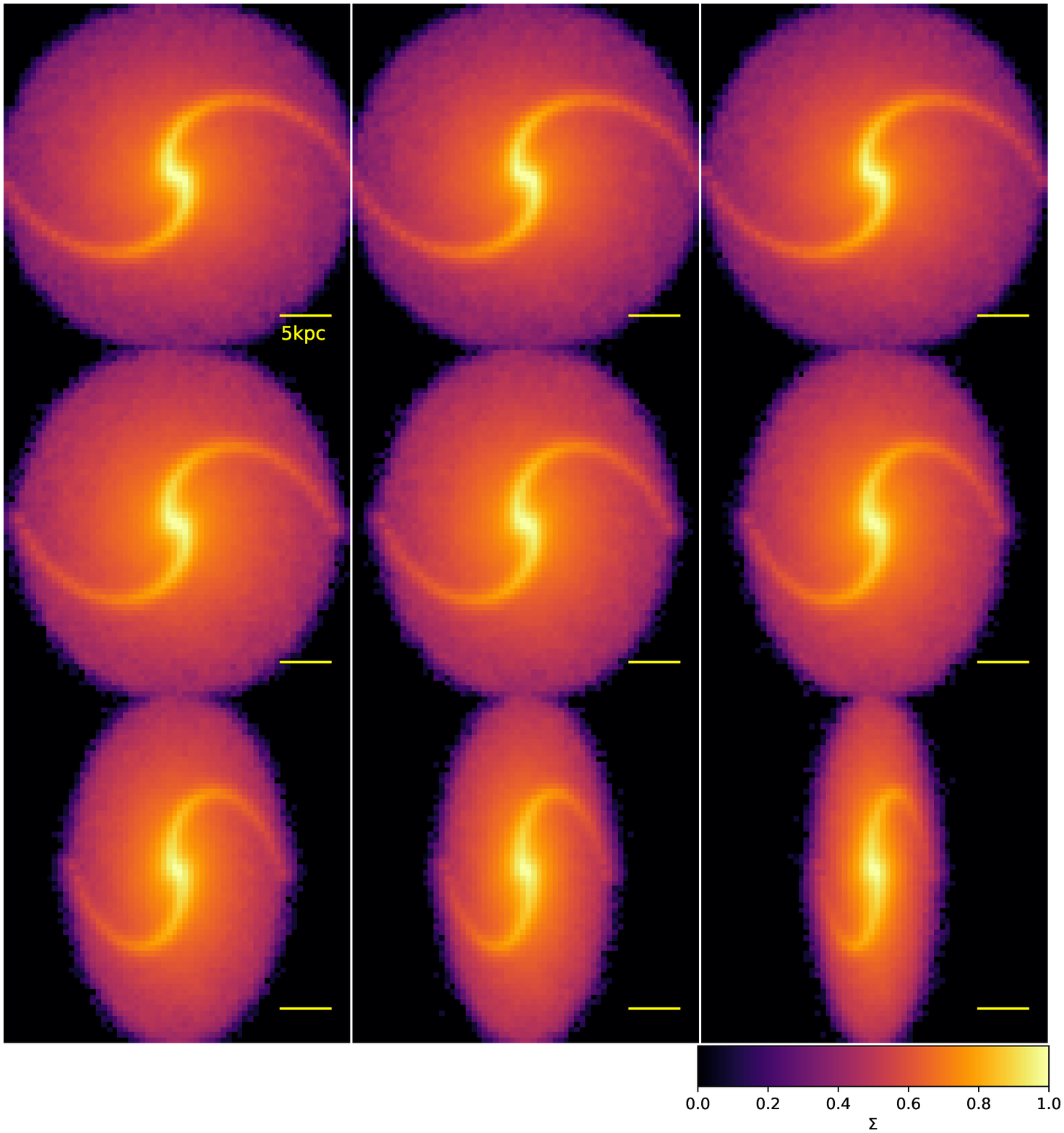}
   \caption{
Same as Fig. 2, but for nine snapshots of the same spiral galaxy model
with $N=2$  viewed from
nine different angles ($\theta$).
               }
              \label{FigGam}%
    \end{figure*}
\subsection{Normalized mass density maps}

The input image data of a spiral galaxy
are  the surface mass density of stars ($\Sigma$ on a logarithmic scale) 
in this study.
We generated galaxy images in the same way as we did in our 
previous works to apply deep learning to galactic mass
distributions (Bekki et al. 2019; Bekki 2019; Shen \& Bekki 2020).
The method is described in Bekki (2019), and we here briefly describe it.
We first divide a galaxy into $64 \times 64$ regions (pixels)
and thereby estimate the surface mass density at each pixel using
a Gaussian smoothing kernel with the smoothing length of $0.03R_{\rm d}$.
The image size of an observed real  galaxy
(i.e., the frame size of a post-stamp image) can be significantly
larger than the apparent size of the galaxy.  We accordingly consider that
the size ratio of the image to the disk is a free parameter represented by
$R_{\rm f}$. Here, we mainly investigate the models with $R_{\rm f}=1$
(where $R_{\rm d}$ is the disk size and fixed at 1 for convenience).
Each 2D $\Sigma$ map has $64 \times 64$ pixels 
and $\Sigma$ is normalized using all pixel values
so that the minimum and maximum
density can be 0 and 1, respectively.

The pixel size was $0.03 R_{\rm d}$ for $R_{\rm f}=1,$
which corresponds roughly to 0.5 kpc for
a Milky Way-type disk galaxy.
The location of each pixel in an image is specified by $i$ ($x$-direction;
$1\le i \le 64$)
and $j$ ($y$-direction; $1 \le j \le 64$), and 
the normalized  2D density  map at $(x_i, y_j)$ can thus  be derived  as follows:
\begin{equation}
\Sigma_{i,j}^{\prime}  =  \frac{ \Sigma_{i,j} - \Sigma_{\rm min}  }
{ \Sigma_{\rm max} - \Sigma_{\rm min} },
\end{equation}
where $\Sigma_{\rm min}$ and $\Sigma_{\rm max}$ are the minimum and maximum values
of $\Sigma$, respectively,
 among the $64 \times 64$ pixels in a model for a given projection.
Therefore the normalization factor is different in different
spiral galaxy  models with
different spiral arm properties viewed from  different projections. 
A spiral galaxy is viewed from 100 different angles in the preset study.

\subsection{Prediction accuracy evaluation}

We investigated how accurately the adopted U-Net can predict the locations of 
spiral arms for each synthesized image using the following metrics. 
Each pixel in a galaxy image was given a component identifier, $I$,
which was 1 or 0 depending on whether it was part of spiral region
(1) or nonspiral region (0). When this $I$ for a pixel of the predicted image 
was the same as $I$ for the same  pixel of 
the original (ground truth) image, then the prediction
accuracy indicator, $P$, was given 1 for the pixel, otherwise, $P$ was given 0. 
the mean $F$ ($F_i$) of $i$th image is given as follows:
\begin{equation}
F_i = \frac{1}{\rm n_{p}} \sum_{\rm k=1}^{\rm n_{p}} P_{\rm k},
\end{equation}
where ${\rm n}_{\rm p}$ is the total number of pixels ($=64 \times 64$  = 4096) 
in an image and $P_{\rm k}$ is the $P$ value at $k$-th pixel.
It should be noted here that this is similar to the standard F-score but still different.
When we assume that spiral and nonspiral regions are
defined as positive (P) and negative (N), respectively,
then spiral and nonspiral  regions that are classified
as spiral regions are labeled TP (true positive) and FP (false positive),
respectively,
in the standard F-score,  
whereas spiral and nonspiral  regions that are classified
as nonspiral regions are labeled FN (false negative) and TN (true negative),
respectively.
We did not use the F-score that is defined as 
$\frac{ \rm TP }{\rm TP + 0.5 \times (FP  +FN) }$
because we need to select both spiral and nonspiral regions that
are correctly segmented

We also separately investigated the prediction accuracy  for spiral arm ($F_{i, \rm sp}$) and
nonspiral arm regions ($F_{i, \rm nsp}$) for each $i$th image
in order to discuss which of the two regions
U-Net can predict $P=1$  more often.
By definition, 
\begin{equation}
F_{i, \rm sp} = \frac{ \rm N(TP) }{ \rm N(TP) + N(FN) },
\end{equation}
where N(TP) and N(FN) are the number of pixels labeled TP and FN, respectively:
this corresponds to a recall rate.
Likewise,
\begin{equation}
F_{i, \rm nsp} = \frac{ \rm N(TN) }{ \rm N(TN) + N(FP) },
\end{equation}
where N(TN) and N(FP) are the number of pixels labeled TN and FP, respectively.

In these prediction accuracy measurements, 
higher $F_i$ in an image  means that the locations of spiral
and nonspiral regions
are more accurately predicted for the image.
We also estimated the average value of these $F$  
($F_{\rm m}$)  for each spiral galaxy model set
as follows:
\begin{equation}
F_{\rm m} = \frac{1}{\rm N_{m}} \sum_{\rm i=1}^{\rm N_{m}} F_i,
\end{equation}
where $N_{\rm m}$ is the total number of images per model set (fixed at 1000).
As done for $F_{\rm m}$, the averages of these
($F_{\rm m, sp}$ and $F_{\rm m, nsp}$) were estimated
from all images in a set of models.

\subsection{Training and testing U-Net}

We trained U-Net using six different training data sets. We first used only 
1000 synthesized images of nonbarred spiral galaxies with fixed $N=2$
and $w_{\rm sp}$  and
variable $f_{\rm sp}$,  $\theta_{\rm max}$, and $f_{\rm bul}$ to train U-Net.
This U-Net and these spiral galaxy models
are referred to as U1 and U1 models, respectively.
Then we trained U2 using 5000 images of nonbarred galaxies
with fixed $w_{\rm sp}$ and variable  $N$, $f_{\rm sp}$,
$\theta_{\rm max}$, and $f_{\rm bul}$. These U1 and U2 models mimicked
late-type spiral galaxies with low bulge-to-disk ratios and different arm properties.
We also trained U3, U4, and U5 by adding large bulges, wider spiral arms, central bars,
and multiple arms with $N=[6-8]$ to the galaxy models used in U2 models. Clearly,
these U-Nets predict the location of spiral arms better for disk galaxies
with different Hubble types because the training data sets themselves contain
images of different Hubble types.
The model parameters for these six model sets are summarized in Table 2.

We tested U-Net in the following two different ways. One 
way was the standard test in which
(i) original images were 
divided into 80\%  and 20\% data sets,
(ii) U-Net was trained using the 80\% data set
and (iii) the trained U-Net was tested using the remaining 20\% data set. 
The other test, which is much more important in this study,
is that U-Net was tested using galaxy images synthesized from
galaxy models that were not used to synthesize galaxy images on which U-Net was trained. In the first test, the training and testing images
were synthesized from the same set of galaxy models. However, in the second
test,  galaxy models from which images were generated on which U-Net was trained are
differed from those for testing U-Net. We focus more on this
second test because the adopted ranges of galaxy models  in this
study should be limited: we cannot model all types of spiral galaxies
with different spiral properties using a limited number of images (that
should be put into U-Net).
If the performance of U-Net is quite good in the second test, then
we can consider that it is very promising in the segmentation of 
spirals arms in disk galaxies with different Hubble types.

\subsection{Ranges of model parameters}

We investigated a large number of spiral galaxy models  with different properties of 
spiral arms, bars, bulges, and inner rings. We considered that the total mass of a disk
galaxy ($M_{\rm gal}$) is fixed at 1 for convenience in data processing  as follows:
\begin{equation}
M_{\rm gal} = M_{\rm bul} + M_{\rm d} + M_{\rm sp} + M_{\rm bar}+M_{\rm ring},
\end{equation}
where $M_{\rm d}$ is the mass of the exponential disk (excluding 
bars, spirals, and rings in the disk).  Therefore
\begin{equation}
1 = f_{\rm bul} + f_{\rm d} + f_{\rm sp} + f_{\rm bar}+f_{\rm ring}.\end{equation}
Each model was allocated the values of these parameters for  given  parameter ranges
(e.g., $0.4 \le f_{\rm sp} \le 0.9$) using a random number generator.

The most important parameter in this study is $f_{\rm sp}$,
as discussed below.  Following the 
results of recent numerical simulations of spiral galaxy evolution by 
Baba et al. (2013),
we assumed a wide range of $f_{\rm sp}$. Although Baba et al. (2013) 
demonstrated 
that $f_{\rm sp}$ (corresponding to their spiral amplitudes) ranges from 0.2 to 0.8
depending on the pitch angles,
we investigated a wider range ($0<f_{\rm sp}<0.9$)  than their predictions, mainly because the observed spiral arms appear to 
show a wider variety than the simulated
ones.

We investigated a large number of testing models  with different 
$f_{\rm sp}$, $N$, $w_{\rm sp}$,
$\theta_{\rm max}$ (controlling pitch angles), 
$f_{\rm bul}$, $f_{\rm bar}$, $R_{\rm bar}$,
$R_{\rm ring}$, $A_{\rm bar}$, 
and $R_{\rm f}$.
Table 3 lists the values of these parameters for 23 testing data sets.
Although wide ranges of these model parameters are explored in this study,
the synthesized images of
spiral arms, bars, and rings were all assumed to be symmetric, which is not realistic when compared to the observed spiral arms. 
These idealized models for spiral galaxies were adopted so that the capabilities
of U-Net in the segmentation of spiral arms could be more quantitatively assessed.
We will discuss the U-Net capabilities in this segmentation task using
more realistic spiral arms models in our forthcoming papers
if U-Net shows a very good performance for these idealized spiral models.

   \begin{figure}
   \centering
   \includegraphics[width=8.5cm]{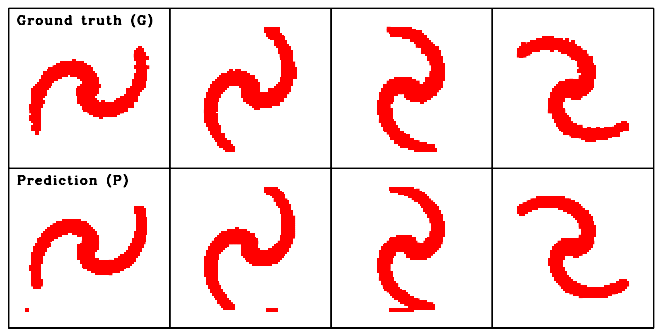}
      \caption{
2D maps of two spiral arms in the original images (ground truth,
or G; upper panel) and
the predicted  images of the arms from U-Net (P; lower panel) for the U1 models.
These images are for $\theta=0$ and different $\phi$.
              }
         \label{FigVibStab}
   \end{figure}
 
   \begin{figure}
   \centering

   \includegraphics[width=8cm]{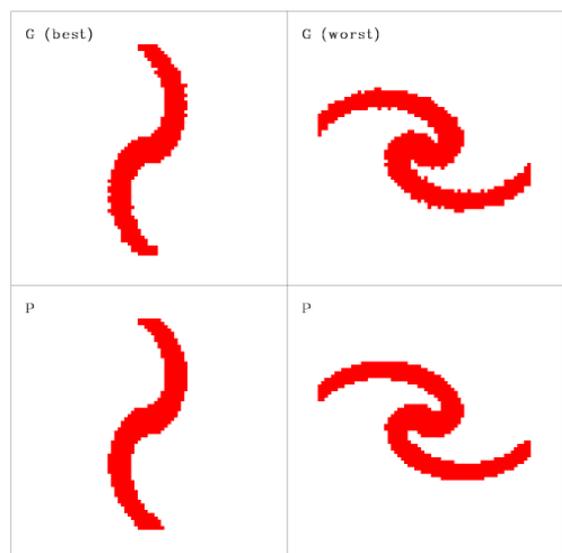}
      \caption{
Same as Fig. 4, but for the two sets of
 best (left) and worst images (right) for which
the prediction accuracy ($F_{\rm m}$) by U-Net is highest and lowest, respectively.
              }
         \label{FigVibStab}
   \end{figure}

\section{Results}

\subsection{Nonbarred models}

Fig. 2 shows six examples of 2D density maps of spiral galaxies
with $N=[1-5]$ and a barred spiral galaxy with $N=2$ viewed
face-on. These clear images with symmetric spiral arms and no significant
inhomogeneity in the exponential disks are input data sets
(training data) that were used to train the adopted U-Net
in this study, although these images are not realistic.
The initial stellar disk in a spiral galaxy model was assumed to have
its spin axis aligned with the $z$-axis, and 
100 images were generated for different view angles, $\theta$ and $\phi$,
where $\theta$ is the angle between the $z$-axis and the line of sight,
and $\phi$ is the azimuthal angle measured from the $x$-axis of the projection
of the line of sight.
Fig. 3 shows nine example images with different $\theta$ for a fixed $\phi$ (=0 degree)
in a non-barred spiral galaxy model with $N=2$.
In order to avoid almost edge-on views of disk galaxies,
we considered that $\theta$ should be lower than 80 degrees.
Using these images with different $\theta$ and $\phi$, we
assessed the capability of U-Net in spiral arm segmentation tasks.

Fig. 4 demonstrates that the predicted (P)
images of spiral arms are very similar to the ground truth (G) images in 
these  U1 models with $N=2$,
which means that the U-Net model U1
can predict the locations of spiral arms very accurately.
However, nonspiral regions close to the edges
of the two arms are identified as arm regions in the three images. This misidentification
(i.e., incorrect prediction)
of arm regions can be seen in all models, although the probability of this misidentification
is rather low ($<0.02$ for most models) in this study.
The mean accuracy ($F_{\rm m}$) for this model is 0.987, which means that
the prediction accuracy of U1 is very good.
Fig. 5 shows that the level of differences in the 2D $\Sigma$ maps 
between predicted (segmented) spiral arms  and ground truth
ones in the best predicted image
is similar to that for the worst predicted images,
which again confirms the very good prediction accuracy of this U1.
The worst predicted image is selected, mainly because the true
spiral regions in the original (ground truth) 
image are not as well identified as such in the predicted image.

Fig. 6 clearly demonstrates that nonspiral regions in original images
are more accurately predicted ($F_{\rm m}=0.995$) than spiral regions  
(0.939) in this model
with $N=2$. The edges of two spiral arms are more likely to be misidentified as
nonspiral regions by U-Net, although the number of such misidentified pixels is small 
(only $<100$ of 1638 spiral regions  for $f_{\rm sp}=0.4$).
In this particular $N=2$ late-type  spiral model,
the prediction accuracy ($F_{\rm m}$) does not depend on
the two inclination angles, $f_{\rm sp}$, $\theta_{\rm max}$,
and $f_{\rm bul}$, although the total number of training and testing images 
is small ($1000$).
The lack of $\theta$-dependence in this spiral segmentation based on U-Net
means that there is no need to deproject galaxy images for this
U-Net.  This would be a remarkable
 advantage of this U-Net because other methods need to deproject
galaxy images by assuming disk inclination and position angles.
   \begin{figure}
   \centering
   \includegraphics[width=8.5cm]{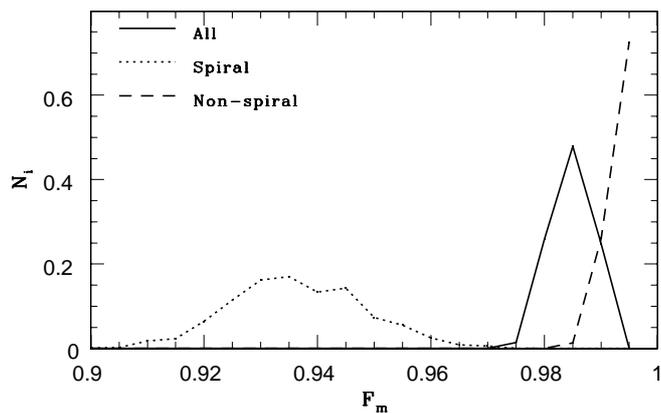}
      \caption{
Number distributions of images
for all ($F_{\rm m}$; solid), spiral arm regions ($F_{\rm m, sp}$; dotted),
and nonspiral arm regions ($F_{\rm m, nsp}$; dashed) for the U1 models.
The normalized number ($N_{\rm i}$) of images for each $F_{\rm m}$ bin is shown.
              }
         \label{FigVibStab}
   \end{figure}

   \begin{figure}
   \centering
   \includegraphics[width=8.5cm]{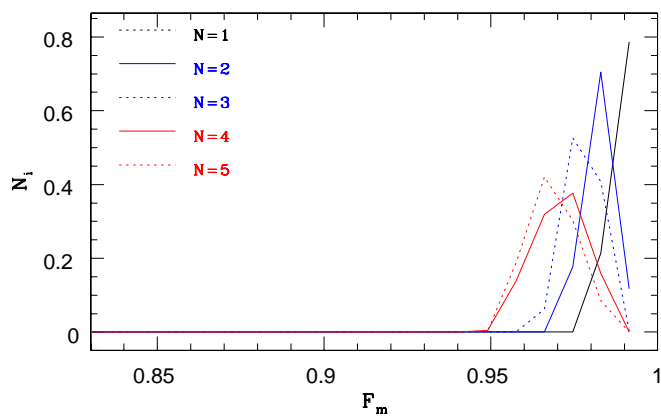}
      \caption{
Same as Fig. 6, but for all regions ($F_{\rm m}$)
in  the U2 models with $N=1$ (dotted black), 2 (solid blue),
3 (dotted blue), 4 (solid red), and 5 (dotted red).
              }
         \label{FigVibStab}
   \end{figure}

   \begin{figure}
   \centering
   \includegraphics[width=8.5cm]{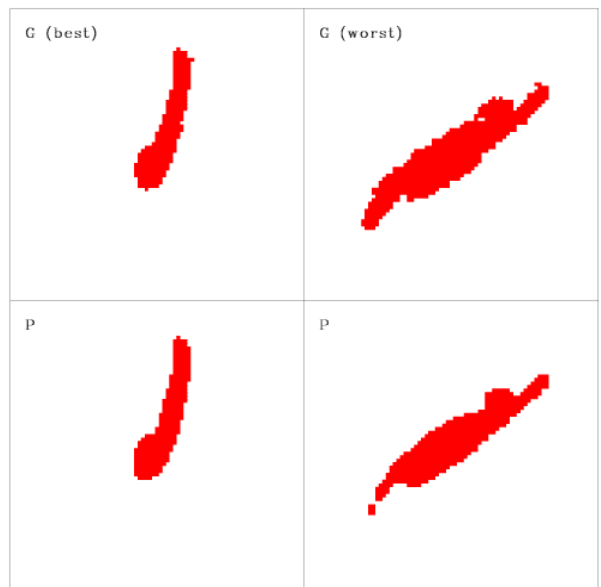}
      \caption{
Same as Fig. 5, but for the U2 models.
              }
         \label{FigVibStab}
   \end{figure}

As shown in Fig. 7,  the prediction accuracy of U-Net trained and tested by 5000 images with
different $N$ (U2) is rather high ($0.982$) for all testing images with different $N$:
$F_{\rm m}$ distributions peak at $F_{\rm m}>0.96$.
This U2 tends to predict less accurately (i.e., smaller $F_{\rm m}$)
for testing 
images with larger $N,$ and the $F_{\rm m}$ distributions are likely to be broader
for these images with larger $N$.
The derived weak dependence on $N$ strongly suggests that as long as a large
number of images with different $N$ are used to train U-Net, it can show
great performance regardless of $N$.

Fig. 8  confirms that the 2D $\Sigma$ maps of segmented spirals
are very similar between the worst predicted
image ($F_{\rm m}=0.93$)
and the original. For the highly inclined disk with two spiral arms, it is
slightly difficult for U2 to predict the locations of the two arms in the worst
predicted image. As expected,  for disks with one arm, it is easier for U2 to
predict the location of the one arm in spiral galaxies owing to its simpler structure:
in this best predicted case, $F_{\rm m}=0.998$.
Fig. 9 shows that nonspiral regions can be more accurately located by U2
than spiral regions, which is seen in U1 models as well.
This means that the probability of spiral regions to be misclassified
as nonspiral is higher than that of nonspiral regions to be misclassified
as spirals.
The peak locations of $F_{\rm m}$
distributions for spiral and nonspiral regions in U2 models 
are almost the same as those seen in  U1 models.
Figure 10 demonstrates that the prediction accuracy of our U-Net
does not depend strongly on the adopted $\theta$ range.

When the spiral arms are segmented, their properties such as pitch angles, total masses,
and arm numbers can be derived straightforwardly (e.g., Davis \& Haynes 2014).
It should be stressed here that this segmentation process by U-Net can be done very
quickly even without using GPU. For example, the segmentation of 1000 testing images by U2 
required only 40 seconds (i.e., 1500 images per minute) for a 
a Linux machine (Centos ver. 7.4) with an Intel i5-4570 CPU at
3.2 GHz frequency (which is not the latest CPU). Accordingly, roughly $2 \times 10^6$ 
disk galaxy images
can be processed and segmented by U-Net in one day.
This means that when GPU machines are used for segmentation tasks by U-Net, then
about $10^7$ images can be segmented in one day by one standard GPU machine.
Recently, Hewitt \& Treuthardt (2020) estimated pitch angles of spiral arms 
using a new parallelized 2D fast Fourier transform algorithm (p2DFFT)
and compared the new method with others.
These standard FFT-based  methods have been used to derive spiral properties, but the  high speed of U-Net
suggests that spiral arm properties can be derived
from spiral arms segmented by  U-Net very quickly as well.

   \begin{figure}
   \centering
   \includegraphics[width=8.5cm]{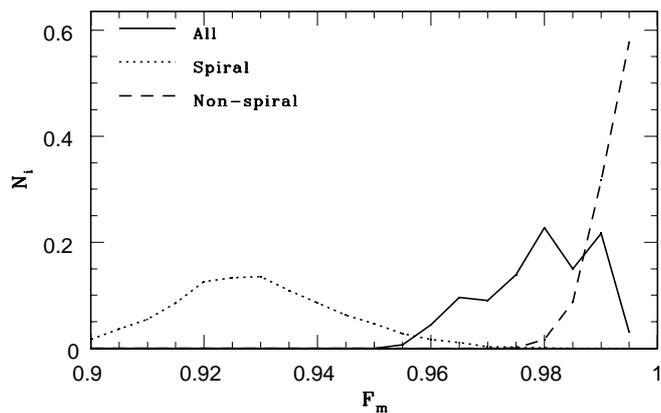}
      \caption{
 Same as Fig. 6, but for the U2 models.
              }
         \label{FigVibStab}
   \end{figure}

   \begin{figure}
   \centering
   \includegraphics[width=8.5cm]{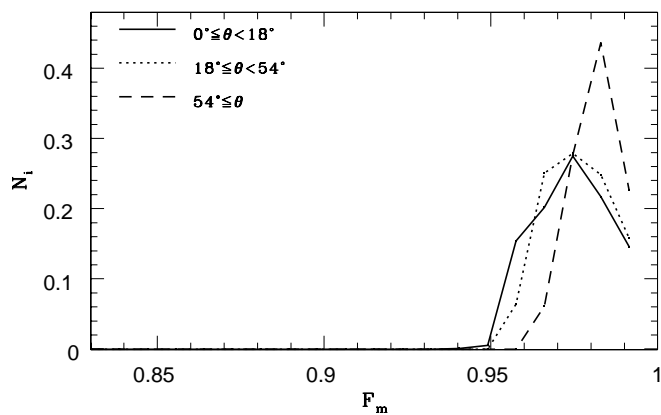}
      \caption{
Same as Fig. 6, but for three different $\theta$ (inclination
of disks) ranges in the U2 models.
              }
         \label{FigVibStab}
   \end{figure}

\subsection{Effects of stellar bars}

The observed spiral galaxies show a much wider variety of spiral
arm structures than the synthesized images of spiral galaxies
in this study. 
Although the original 5000 images in  U2 models  were split into 4000 to train U-Net
and 1000 to test it,  these images were synthesized from disk galaxies with
certain ranges of model parameters (e.g., $N=[1,5]$ and $f_{\rm bul}<0.1$).
Therefore it is not clear whether U-Net
can predict the locations of spiral arms in  galaxies with $N \ge 6$ or $f_{\rm bul} \ge 0.2$.
If  U2
can very accurately
 predict the locations of spiral arms in images of spiral  galaxies that
are not used to generate synthesized images to train U1 
(i.e., entirely unknown data sets), then we can consider that U-Net
is a powerful tool for the segmentation of spiral arms.
Thus we here tested U1 using entirely unknown data sets.

The central stellar bars in  spiral galaxies with $N=2$ would make it more 
difficult for U-Net to predict the locations of the two arms because
the two arms are connected to the similarly bisymmetric bars.
Fig. 11 compares the 2D distributions
of original and predicted (segmented) spiral arms for four 
example images in T10 testing  models. 
As expected, the parts of the inner stellar bars are  misclassified as the parts
of two spiral arms in these images, although $F_{\rm m}$ is still high
in these images ($F_{\rm m}=0.958$ for the testing model T10). 
In the image of a highly inclined disk,
the entire bar regions are  misclassified as spirals, which would be
a serious disadvantage of U-Net in the segmentation of spiral arms.
The second image from the left in Fig. 11 shows nonspiral regions that
are misclassified as spirals, although other images do not show this type
of misclassification. 

We confirm that spiral arms in barred spiral galaxies with different $N$
but a fixed $f_{\rm bul}$
can be precisely segmented by U2 with $F_{\rm m}>0.95$ (see Table 3 for
T10, T11, and T12 models). Furthermore,
we found that  
bar sizes and shapes do not affect the prediction accuracy of U2
for spiral galaxies with a fixed $f_{\rm bul}$. For example, as shown 
in Fig. 12, U2 can precisely segment spiral arms for the two barred  
galaxies with $R_{\rm bar}=0.2$ and 0.5, although their $F_{\rm bar}$
distributions are slightly different for a narrow range of $F_{\rm bar}$
($>0.85$). Spiral arms of barred spiral galaxies 
with shorter bars are more likely to
be precisely segmented in U2.
These results imply that U-Net can be applied to barred and nonbarred
spiral galaxies, although the stellar bar regions of barred spiral galaxies
viewed almost edge-on can be misclassified as spiral arm regions
by U-net. This misclassification problem of U-Net will need to be
addressed in our future works using more complicated architectures of U-Net
and a greater variety of spiral arm structures.

   \begin{figure}
   \centering
   \includegraphics[width=8.5cm]{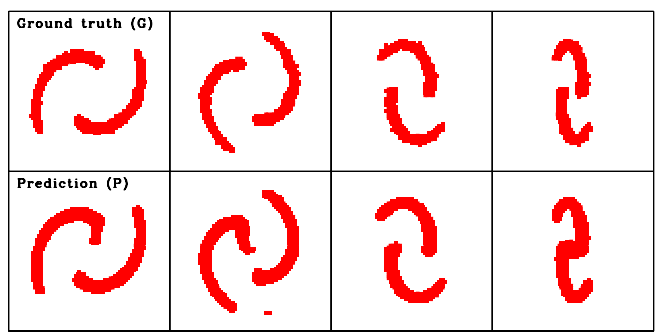}
      \caption{
 Same as Fig. 4, but for testing images ($T10$)  of barred spiral galaxies
with $N=2$. This segmentation was made by U2, which was trained on synthesized images
of nonbarred galaxies. Nevertheless, the locations of segmented spirals appears to
match those of the original (ground-truth) spirals well.
              }
         \label{FigVibStab}
   \end{figure}

\subsection{Parameter dependence}

It is clear from Fig. 12 that bulge mass fractions do not affect the 
prediction accuracy of U-Net in the segmentation of spiral arms. 
It should be stressed here that  these
images were synthesized from large bulge models with $0.1<f_{\rm bul}<0.5$ that
were not used in the training phase of U2.
Accordingly, this result suggests that
U-Net can accurately predict the locations of spiral arms for
disk galaxies with a wide variety of bulge types
with different $f_{\rm bul}$, even if the U-Net is trained 
by galaxy images with a limited range of $f_{\rm bul}$. This is promising
because it would be unrealistic to model all different galaxy morphologies in the
universe
using a huge number of galaxy images for these.
Fig. 12 also demonstrates that spiral arms in galaxy images with stronger arms
can be more accurately segmented by this U2 in large bulge models.

As shown in Fig. 12, $F_{\rm m}$ depends on $w_{\rm sp}$ such that  it is
lower for quite wide spiral arms with $w_{\rm sp}=0.06$ (T13), corresponding to 30\% of the
disk scale length. Although the locations of spiral arms in these images
with wide spiral arms are least accurately predicted, the mean $F_{\rm m}$ is
still relatively good ($\sim 0.9$ for T13; see Table 3).
 We confirm that if galaxy images with wide
spiral arms are included in the training data set for U-Net, then the trained
U-Net can have a better prediction accuracy for these wide spiral arms
($F_{\rm m}=0.952$; see Table 2 for U5).

Although the shapes and the 
numbers of spiral arms in disk galaxies are observed to be widely 
different (e.g, Iye et al. 2019; Tadaki et al. 2020),
$\sim 90$\%  of these have $N\le 5$ (e.g., Hart et al. 2017).
It is, however, 
an important test whether U-Net trained by images with a limited range
of $N$ in U2 models can predict the locations of spiral arms in galaxies with 
$N>5$.
Fig. 12 shows that (i) the prediction accuracies for images with $N=6$ and 7
are good ($F_{\rm m}>0.93$)
and (ii) the $F_{\rm m}$ distributions are wider than those for $N=[1-5]$.
The high accuracies obtained for $N=6$ and 7 are surprising,
given that
the range of  spiral arm numbers in the training data sets was limited in U2 models
($N=[1-5]$).
These results again imply that a huge number of images with a large variety of $N$ 
(e.g., $N=[1-10]$ are
not required to develop U-Net that can segment spiral arms accurately for
disk galaxies with different $N$. Because the observed shapes of flocculent spiral 
arms are not modeled in this study, it remains to be clarified whether
U-net trained on galaxy images with symmetric spiral arms
can accurately segment such flocculent arms.

   \begin{figure}
   \centering
   \includegraphics[width=8.5cm]{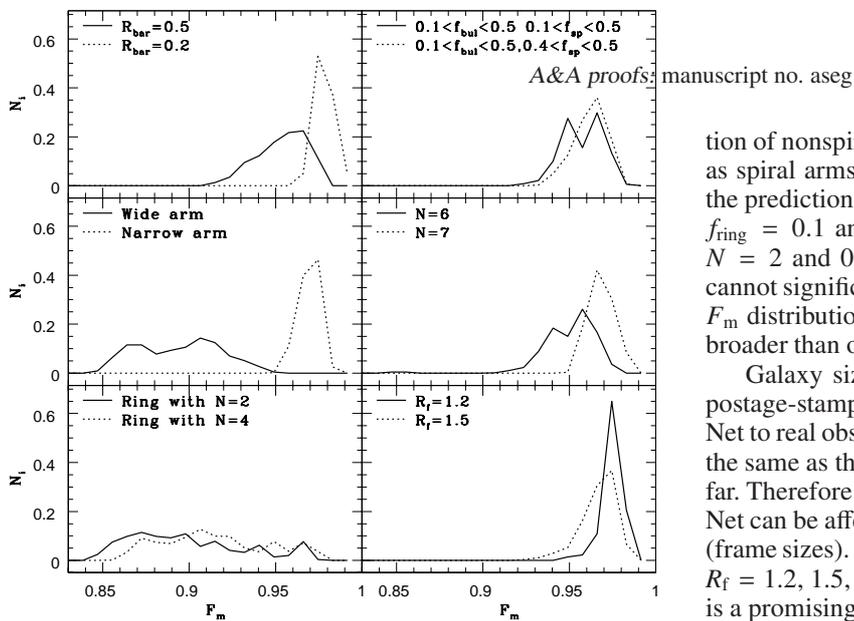}
      \caption{
Same as Fig. 6, but for 12 different testing data sets: different $R_{\rm bar}$
(T10 and T12; top left), $f_{\rm bul}$ (T1 and T2; top right),
$w_{\rm sp}$ (T13 and T14; middle left), $N$ (T15 and T16; middle right),
ring models with different $N$ (T17 and T18; bottom left),
and different $R_{\rm f}$ (T21 and T22, bottom right).
The U-Net used for these tests was trained on
disk models that were not used to generate these images for testing.
              }
         \label{FigVibStab}
   \end{figure}

We found that U2 can accurately segment spiral arms with $F_{\rm m}>0.95$ only
if $f_{\rm sp}>0.2$, that is, only when disk galaxies have stronger arms (see Table 3
for T3-T7).
The prediction accuracy depends on $f_{\rm sp}$ in such a way that
$F_{\rm m}$ is higher for larger $f_{\rm sp}$: $F_{\rm m}=0.761$,   0.809,
0.887, and 0.954 for $f_{\rm sp}$=[0-0.05], [0.05-0.1],  [0.1-0.2], and [0.2-0.3],
respectively. These results imply that disk galaxies with weaker spiral arms,
such as the so-called anemic spirals in the Virgo galaxy cluster, would not be able to
be segmented well by U-Net. However, the prediction accuracy of 0.76 for
very weak spiral arms is encouraging: still 76\% of the spiral or nonspiral regions 
can be accurately located by U-Net.

Inner rings around stellar bars in  barred spiral galaxies are found to slightly affect the prediction accuracy in our models with
$N=2$ and 4,
$f_{\rm ring}=0.1$,  and $R_{\rm ring}=R_{\rm bar}=0.2$ and 0.5.
For example, $F_{\rm m}$ in the barred spiral models with
a ring with $R_{\rm ring}=0.5$ and $N=2$ (4) are  0.903 (0.919),
which is slightly lower than the similar model without rings (see Table 3 for T17 and T18).
However, it should be noted here that $F_{\rm m, sp}$ and $F_{\rm m, nsp}$
in this barred spiral model with $N=2$ and a ring
are 0.982 and 0.891, respectively, which means that spiral regions are more 
accurately predicted by U2 than nonspiral regions.
A minor fraction of nonspiral regions including the inner ring is misclassified
as spiral arms, which means that inner rings can slightly lower the prediction
accuracy of U-Net.
The barred spiral models with $f_{\rm ring}=0.1$ and
$R_{\rm ring}=0.2$ show rather high $F_{\rm m}$ of  0.976 for $N=2$ and 0.972 for $N=4$,
which means that smaller rings cannot significantly affect the prediction
accuracy of U-Net.
The $F_{\rm m}$ distributions for these two testing data sets are significantly
broader than other models sets.

Galaxy sizes can be 
significantly smaller than the  sizes of postage-stamp images, which 
we need to use when we apply U-Net to real observations.
However, 
the sizes of testing images  are the same as those
of galactic disks
in all of the models shown so far.
Therefore we investigated how the prediction accuracy of U-Net 
can be affected by the ratios of galactic disks to images sizes (frame sizes). Fig. 12 confirms that $F_{\rm m}$ is rather high ($>0.97$)
for $R_{\rm f}=1.2$, 1.5, and 2.0  (see Table 3 for T21, T22, and T23).
This  is a promising result because we do not need to adjust the image sizes so that
they can be similar to galactic disk sizes (e.g., by cutting the images properly).

This shows that even though these testing images
were synthesized by using disk galaxy models that were not used to generate the training
data sets for U2,
$F_{\rm m}$ of U2 are overall good for these testing images.
Furthermore, U3-U5, which were trained using different sets of disk galaxy
models, show very good prediction accuracies ($F_{\rm m}>0.98$: see Table 2).
Although these results are all promising, it is a fundamental question
whether U-Net trained by a limited number of images from a limited
number of disk galaxy models can really segment all different types
of spiral arms in disk galaxies with different Hubble types in a very
accurate manner. This question will be addressed when several features
of spiral arms that were not modeled in this study 
(e.g., flocculent shapes, nonsymmetry, nonconstant pitch angles,
and unequal spacing of arms)
are properly considered in generating spiral arms in disk galaxies.

   \begin{figure*}
   \centering
  \includegraphics[width=18cm]{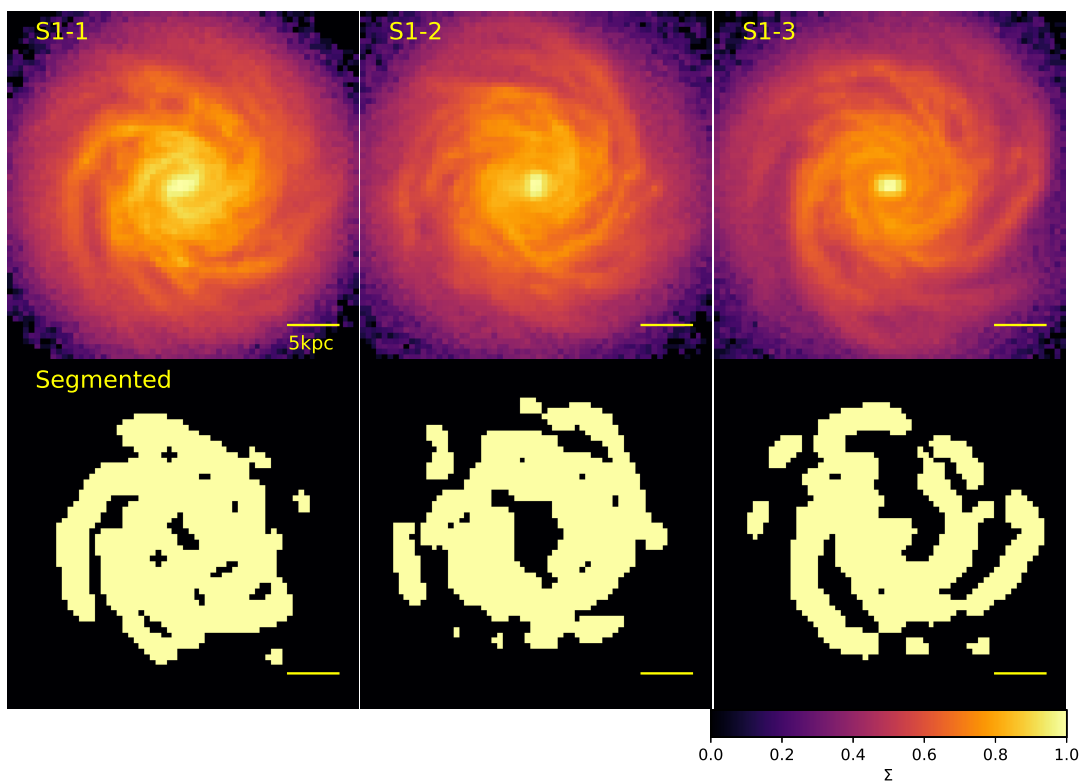}
   \caption{
Two-dimensional density maps of the spiral galaxy model S1 for three different time steps
(S1-1, S1-2, and S1-3; upper panel) and 
the projected distributions of the segmented spiral
arms (lower panel). For clarity, the pixels that are identified
as the parts of spiral arms are shown as $\log \Sigma=1$
(i.e., highest density; yellow). Other pixels are shown as 
$\log \Sigma=0$ (i.e., lowest density; black).
               }
              \label{FigGam}%
    \end{figure*}

   \begin{figure*}
   \centering
  \includegraphics[width=18cm]{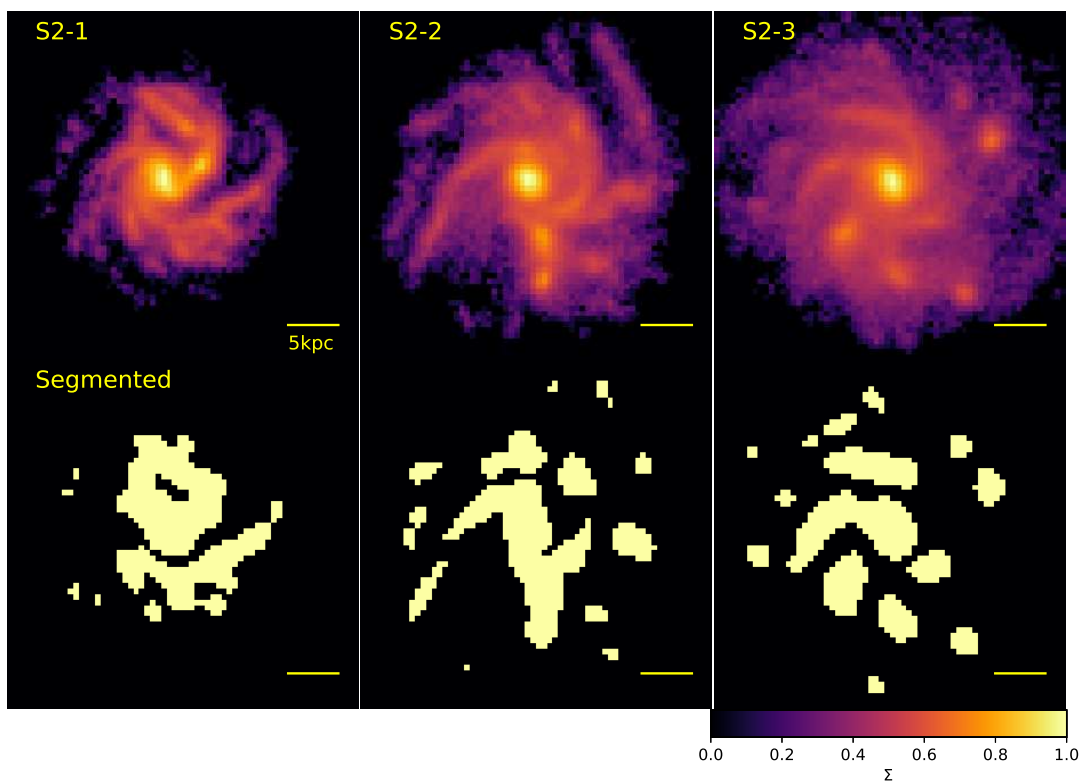}
   \caption{
Same as Fig. 13, but for the S2 model (S2-1, S2-2, and S2-3).
               }
              \label{FigGam}%
    \end{figure*}

   \begin{figure*}
   \centering
  \includegraphics[width=18cm]{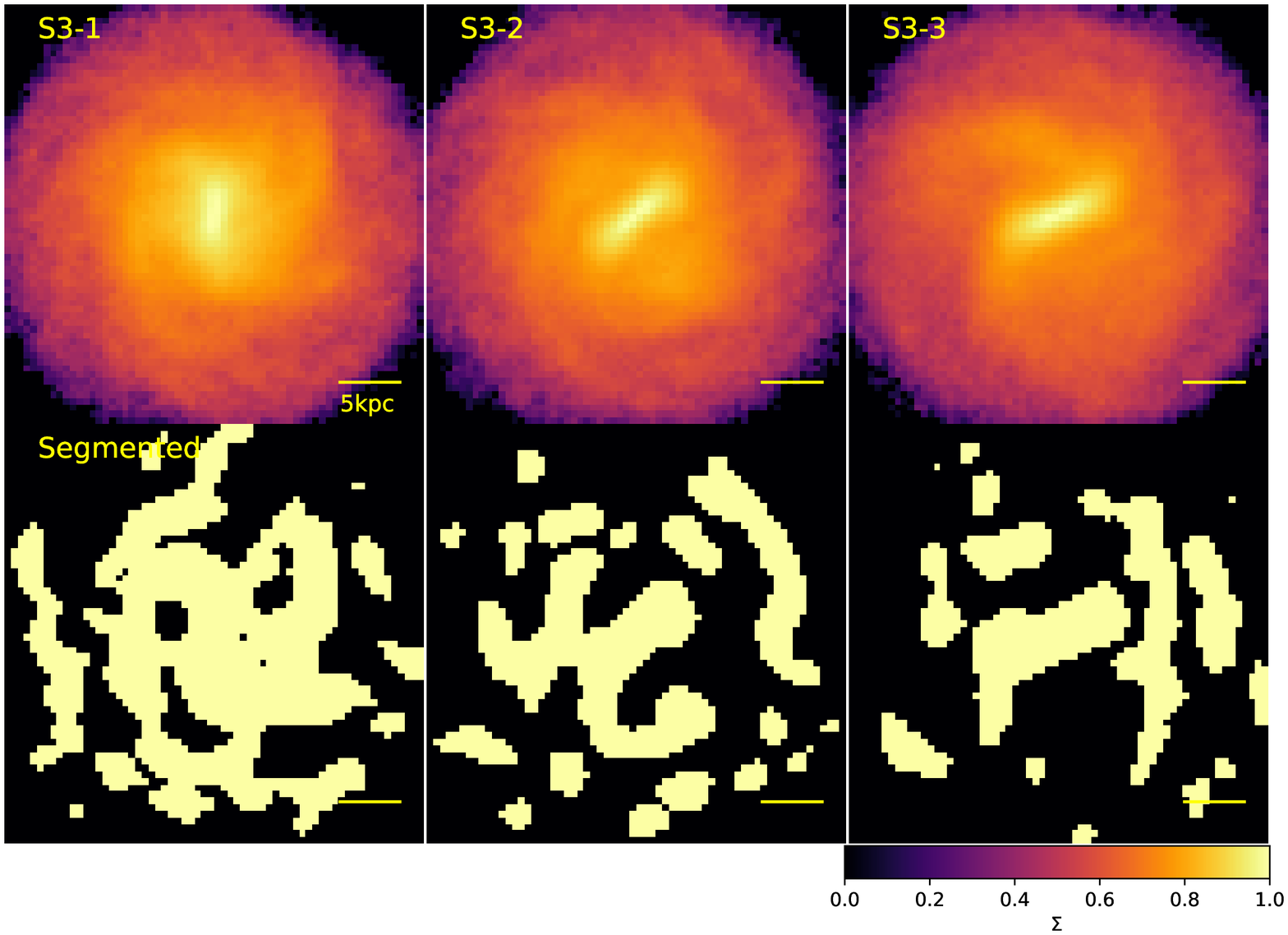}
   \caption{
Same as Fig. 13, but for the S3 model (S3-1, S3-2, and S3-3).
               }
              \label{FigGam}%
    \end{figure*}

\section{Discussion}

\subsection{Tests using hydrodynamical simulations of disk galaxies}

The present study is based on many images of spiral galaxies that are based on mathematical models (i.e., symmetric and regular spiral arms). Since these images are less realistic compared
to the observed spiral galaxies with asymmetric and irregular spiral arms,  here we discuss
how we improve our segmentation models using more realistic spiral galaxy images.

\subsubsection{Limitations and further improvement}

Although
we have clearly demonstrated that U-Net can 
predict the precise locations of spiral arms in disk galaxies with
different Hubble types,
the following assumptions  in the adopted spiral models
might lead to worse predictions of the locations of spiral
arms in the application of U-Net to real observations.
First, all images we used were assumed to be symmetric, which 
is not realistic. This means that when U-Net is applied to 
spiral galaxies with  different pitch angles between different arms,
then the prediction accuracy can be lower when U-Net is applied to real observed images.
Moreover, when spiral galaxies have lopsided distributions of stars,
U-Net might predict the locations of spiral arms less accurately.
Second, no localized density enhancement along arms was considered
in disk galaxies for simplicity and clarity. 
Formation of massive OB stars in some areas of spiral arms 
can significantly enhance the luminosity densities of the ares
so that the arm can have bright knotty region.
These knotty star-forming regions in the arm can make it more 
difficult for U-Net to identify global spiral arms in disk galaxies.

Third, two spiral arms in a barred spiral
galaxy were assumed to originate from the two edges of the bar.
Although the observed images of barred spiral galaxies
suggest that this assumption is quite reasonable and realistic,
some of these galaxies have 
secondary bars. Because inner  secondary bars
within primary bars were not modeled at all,
U-Net might not predict the locations of spiral arms
in disk galaxies with 
these fine structures.
Fourth,  only logarithmic spiral arms with $N=1-8$ and  different
pitch angles were modeled. Accordingly,
U-Net would not be able to predict the locations of
flocculent spiral arms observed in a significant fraction
of spiral galaxies precisely (Elmegreen 1981).

\subsubsection{Three different types of simulated spiral galaxies}

In order to discuss whether U-Net trained on the present models for
symmetric spiral arms with constant pitch angles can predict the
locations of nonsymmetric and irregular spiral arms with their pitch
angles depending on the arm positions well, we here use the results of
our original hydrodynamical simulations of disk galaxies (Bekki 2013, 2014b).
The code of the simulations
enables us to investigate the evolution of neutral and molecular
hydrogen gas, metals, and dust in disk galaxies
in a fully self-consistent manner.
The details of the code are given in Bekki (2013, 2015, 2017).
We investigated three different spiral galaxies, labeled S1, S2,
and S3. S1 is a gas-rich Milky Way-type luminous disk galaxy 
with a gas mass fraction ($f_{\rm g}$) of 0.1, and this mimics
the present-day gas-rich late-type disk galaxies.
S2 mimics  high-z very gas-rich spiral galaxies with clumpy
gas distributions in which $f_{\rm g}$ was set to be 0.45.
S3 with $f_{\rm g}=0$ mimics  barred galaxies with very week spiral
arms that are being transformed into S0s through disk heating by
spiral arms. 

In these three models, disk galaxies consist
of dark matter and stellar and gaseous disks, and the disk galaxy
models are the same as those adopted in Bekki (2014b).
The dark matter masses
were set to be  $10^{12} {\rm M}_{\odot}$ for all models. The stellar disk
mass is $6 \times 10^{10} {\rm M}_{\odot}$ for S1 and S3
and $3 \times 10^{10} {\rm M}_{\odot}$ for S2.
After 3 Gyr evolution of these models, snapshots at all time steps
were inspected by eye, and accordingly,  three snapshots with
spiral arms were selected for each of the three disk models.
These three images were labeled, for example, S1-1, S1-2, and S1-3
for the S1 model.
The 2D maps of mass distributions were then input into U-Net
model U4 (see Table 2), which was trained on a large number of images
from barred spiral galaxy models.
Figs. 13, 14, and 15 describe the locations of segmented spiral arms
in the $x$-$y$ plane for these nine images.

Clearly, the simulated spiral arms are irregular and much less symmetric,
with pitch angles being different depending on the positions of the arms, 
which are
in a striking contrast with those shown in \S 3.  
Nevertheless, Fig. 13 shows that at least some spiral arms with
high stellar densities are well
segmented by U-Net. 
In S1-1 and S1-2, the central small short bar ( and bulge) is
not misclassified as spiral arms, 
but the central bar in S1-3 is identified as the root of the inner
spiral arm. Because spiral arms are observed to originate from the central
bars in real galaxies, this misidentification problem  would need to be
addressed in our future U-Net with more complicated architectures
for better segmentation processes.
Furthermore, the low-density tips of spiral arms are not well segmented
by U-Net. This is expected because U-Net does not show great performance
in our mathematical spiral galaxy models with weak arms.
The outer spiral arms with low stellar mass densities are likely to be 
missed in this segmentation processed based on U-Net.

As shown in Fig. 14,  clumpy spiral arms are relatively well segmented by U-Net, although the central high-density bulge regions are misidentified as parts
of the arms in these clumpy spiral S3 images. Furthermore, 
massive clumps formed in spiral arms are also identified as parts of
the arms, which appears to be inevitable in this segmentation process.
As a result of this, the spatial
distributions of the  segmented spiral arms  do not look like spirals,
although the original 2D images clearly look like spirals.
It would be a matter of debate whether clumps formed from  spiral arms should be
regarded as parts of spirals.
However, these results for S2 strongly suggest that
we need to improve the performance of U-Net through training U-Net on
a large number of clumpy spiral galaxies.

Fig. 15 shows that
our U-Net can properly identify some of very  weak spiral arms of the barred
spiral galaxy in S3. However, three problems remain
in this segmentation of this S3 model by U-Net.
First, the central strong bar in these three images is misidentified
as a spiral arm: intriguingly, the length and direction of the major
axis of the bar appears to be  well identified.
Second, apparently nonspiral arm regions are classified as spiral arms
in the outer part of the disk. Third, only the local high-density
regions of the weak spiral arms are segmented so that the spatial
distribution of the segmented arms cannot be like spiral arms.
These three problems imply that it would be a difficult task for
U-Net to identify very weak and wide spiral arms in barred spiral
galaxies.

Currently, we do not have an accurate measure to quantify the prediction
accuracy of U-Net for these nonsymmetric and irregular spiral arms
with nonconstant pitch angles because we do not have an independent method
for locating spiral arms precisely: making training data sets for such
arms is a formidable task.
In our forthcoming papers, we will (i)
generate a large number of images
with nonsymmetric and flocculent arms
that are not well modeled in this study
using both constrained and
cosmological hydrodynamical simulations,
and (ii) investigate a method for locating spiral arms without
using U-Net to create training data sets for U-Net.
If U-Net is confirmed to show very good performance even for such
realistic spiral arms in our future studies,
we will be able to apply it  to the real observed images of
spiral galaxies.

\subsection{Multiwavelength science cases}

If U-Net can accuracy find the precise locations of spiral arms for a 
large number of spiral galaxies, then new data sets 
would have the following
significant benefits in multiwavelength studies of galaxies.
First,  such new data sets for
spiral galaxies would enable us to investigate the (mass) fractions of
GMCs that are located in arms (or inter-arm) regions of the galaxies
in a statistical manner. Physical correlations between
the mass fractions of GMCs in spiral arms and the mass fractions
(i.e., strengths)  of spiral arms in disk galaxies can be
also investigated  because it is straightforward to estimate
the mass fractions of spiral arms from segmented spiral arms.
This correlation will greatly advance our understanding of GMC
formation and destruction within spiral arms.
Because spatial distributions of molecular hydrogen in nearby galaxies
have already been investigated with ALMA, 
this type of multiwavelength investigation is currently possible.

Second,
star formation efficiencies in arm and inter-arm regions can be 
separately investigated  based on the locations of spiral arms
and the gas masses of neutral (and ${\rm H_2}$)
hydrogen in the arms. Ongoing large HI surveys such as
the WALLABY project (e.g., Koribalski et al. 2020) are currently
investigating HI properties for a large number of nearby galaxies
and will provide data sets for the internal distribution of HI gas
in these galaxies. Accordingly,  star formation rates
(e.g., based on H$\alpha$ distributions) and  HI column densities
in arm and inter-arm regions can be investigated so that
the star formation efficiencies in
these regions can be derived for a large number of galaxy samples.
This statistical investigation
will allow astronomers to discuss
the key question  of how spiral arms induce galaxy-scale star formation in detail.

Third, the rotation curve profiles and the mass fractions
of spiral arms for a large number of disk galaxies will reveal the physical
connection between the baryonic fractions of galaxies and the strengths of
spiral arms. The next-generation large HI surveys will reveal the maximum circular
velocities and thus the dynamical masses (including dark matter)
within optical radii for a large
sample of nearby galaxies. Moreover, the locations of spiral arms derived from
U-Net can be used to estimate the total masses of spiral arms, thus the mass ratios
of spiral arms to stellar disks. The ratios of baryonic masses to dynamical masses
(i.e., the degree of self-gravitation) can affect the details of spiral structures
in disk galaxies (e.g., Seigar et al. 2008). Therefore, the estimation of spiral arm
mass fractions in disk galaxies observed by ongoing large HI surveys will make significant progress in understanding the formation of spiral arms, which is affected
by their dynamical properties.

Finally, U-Net can be applied for segmentation tasks for gaseous spiral arms 
in a huge number of nearby and distant galaxies investigated by the SKA. Although
identification and characterization of spiral arms in disk galaxies have been performed
for optical and near-infrared images of the galaxies
(e.g., Elmegreen \& Elmegreen 1984; Davis et al. 2014; Hart et al. 2017;
Masters et al. 2019), no data base contains the physical properties of gaseous arms (e.g., numbers and shapes) for
a large ($>1000$) number of disk galaxies. 
Segmentation of spiral arm structures for HI images from SKA
will provide a new radio catalog for the details
of gaseous spiral arms in a huge number of galaxies.
A comparative study of gaseous  and stellar  spiral arms 
based on these radio and optical catalogs of disk galaxies will significantly
advance our understanding of the interplay between gaseous and stellar dynamics in
the evolution of spiral arms.

\section{Conclusions}

We have applied U-Net to a large number of synthesized images of spiral galaxies 
($N \sim 50,000$) in order to
assess the capability of the U-Net to accurately identify the locations of spiral arms.
Because this is the first step in this series of papers on galaxy segmentation,
we used disk galaxy models in which viewing angles of disks ($\theta$ and $\phi$),
spiral arm numbers ($N$), strengths ($f_{\rm sp}$), widths ($w_{\rm sp}$),
and pitch angles (controlled by $\theta_{\rm max}$),  bulge mass fractions ($f_{\rm bul}$),
presence or absence of stellar bars and rings, and image sizes
(with respect to disk sizes)  were quite different.
In the present segmentation tasks, the spiral arms and other regions in disk galaxies
were labeled 1 and 0, respectively, so that
the levels of accuracy in the predicted
locations of spiral arms could be well quantified by $F_{\rm m}$,
which is similar to the standard $F$-scores.
Although the adopted models were rather idealized and less realistic in some physical properties
(e.g., symmetric arms and no background stars), they enabled us to understand the advantages of the U-Net
in  galaxy segmentation tasks. 
The principal results are listed below. \\

(1) U-Net can accurately predict the locations of spiral arms in nonbarred
 disk galaxies with 
$1 \le N \le 8$
and $0.1 <f_{\rm sp} < 0.9$. 
The mean accuracy score ($F_{\rm m}$) depends on $N$, but
it is consistently as high as 0.98 (with 1 and 0 being fully accurate and inaccurate, respectively).
U-Net can also  accurately predict
the locations of spiral arms in disk 
galaxies with different bulge-to-disk ratios ($f_{\rm bul}<0.5$) for a given $N$.
Thus these results strongly suggest that U-Net is quite 
useful in galaxy segmentation tasks for disk galaxies with different Hubble
morphological types. \\

(2) The performance of U-Net in these segmentation tasks is very good
for barred spiral galaxies with $N=2$ ($F_{\rm m} \approx 0.95$). 
These results do not depend on the sizes ($R_{\rm bar}$) and the strengths
($f_{\rm bar}$) of the stellar bars, which implies that
the presence or absence of bars does not affect the prediction accuracy of U-Net
in segmentation of spiral arms.
Furthermore,  $F_{\rm m}$ in barred and nonbarred galaxies
is not different between images with different  viewing angles of 
disk galaxies ($\theta
<80$ degrees). These results imply that segmentation of spiral arms and bars
can be possible even for highly inclined disk galaxies.  \\

(3)  It should be stressed, however, that if  U-Net is  trained on galaxy images
with particular ranges of $N$,
and if galaxy images for testing have $N$ outside the $N$ ranges for the training,
then $F_{\rm m}$ becomes slightly lower.
For example, U-Net trained with images with $N=1$, 2, 3, 4 and 5 can only
segment spiral arms in $N=6$ models  with $F_{\rm m}=0.934$.
This implies that a wide range of $N$ should be used to train U-Net when U-Net is applied for real disk galaxies with wide ranges of spiral arm 
properties and if very high $F_{\rm m}$ ($>0.98$) is required. \\

(4) Although stellar rings around bars do not affect 
$F_{\rm m}$
, the mass fractions of spiral arms ($f_{\rm sp}$) can significantly affect
$F_{\rm m}$ for $f_{\rm sp}<0.1$. For example,
$F_{\rm m}$ can be as low as 0.8 for $0.05 < f_{\rm sp} < 0.1$ in nonbarred disk galaxies.
This is not a problem per se, but 
it means that disk galaxies with very weak
spiral arms, such as anemic spiral arms in clusters of galaxies,
would be harder to segment by U-Net. This is the most remarkable
disadvantage in the segmentation of spiral arms by U-Net in this study. \\

(5) We applied the U-Net trained on images with symmetric spiral arms
to the synthesized images of disk galaxies with nonsymmetric and irregular
spiral arms from our original hydrodynamical simulations of the galaxies.
U-Net cannot accurately predict the locations of the spiral arms in the simulated
disk galaxies with more realistic spiral arm properties, in particular, in barred galaxies
with weaker and wider spiral arms, probably because the training data sets
do not include such nonsymmetric and irregular spiral arms. However, it might be intrinsically difficult for U-Net to segment such nonsymmetric spiral arms.
\\

(6) These results suggest that it is worthwhile for our future studies
to train U-Net with a larger number of more realistic galaxy images
with noise, nonsymmetric spirals,  
and different pitch angles between different arms, etc.
It might not be so straightforward to 
generate many pairs of these complicated images and locations
of spiral arms (i.e., labels), but  
high-resolution hydrodynamical
simulations of disk galaxies with different Hubble types
might be useful for this purpose.
When we can confirm that
U-Net can predict the location of spiral arms for these images very accurately,
then U-Net can be readily applied to real observational images of 
spiral galaxies.
Given that $\sim 10^7$ galaxy images can be processed and segmented by just one GPU machine in one day,
segmentation of spiral structures in galaxies will be very useful for the analysis of a huge
number of images from next-generation telescopes such as the LSST and EUCLID.


\section{Acknowledgment}
I (Kenji Bekki; KB) am   grateful to the referee  for  constructive and
useful comments that improved this paper. 
This  research  was  supported  by  the  Australian  government 
through the Australian Research 
Council's Discovery Projects funding scheme (DP170102344).

%

\begin{thebibliography}{}


\item
 Athanassoula, E. 1984, PhR, 114, 319 \\

Baba, J., Saitoh, T., R., Wada, K., 2013, ApJ, 763, 46 \\

Bekki, K., 2013, MNRAS, 432, 2298 \\

Bekki, K., 2014a, MNRAS, 438, 444 \\

Bekki, K., 2014b, MNRAS, 444, 1615  \\

Bekki, K., 2015, MNRAS, 449, 1625 \\

Bekki, K., 2017, MNRAS, 467, 1857 \\

Bekki, K., 2019, MNRAS, 485, 1924 \\

Bekki, K., et al. 2019, Astron. Comput., 28, 100286 \\

Boucaud, A., et al. 2020, MNRAS, 491, 248 \\

Buta, R., 2013, Secular evolution of galaxies, XXIII Canary Istands Winter School of
Astrophysics (arXiv:1304.3529) \\
 
Carlberg, R. G., Freedman, W. L., 1985, ApJ, 298, 486 \\

Cavanagh, M. K., Bekki, K., 2020, A\&A, 641, 77 \\

Chollet F., 2015, Available at:https://keras.io/ \\

Danver, C.-G., 1942, AnLun, 10, 162 \\

Davis, B. L., et al., 2012, ApJS, 199, 33 \\

Davis, B. L., et al., 2017, MNRAS, 471, 2187 \\

Davis, D. R., Hayes, W. B., 2014, ApJ, 790, 87 \\

Diaz, J. D., et al., 2019, MNRAS, 486, 4845 \\

Dieleman, S., Willett, K. W., Dambre, J., 2015, MNRAS, 450, 1441 \\

Dobbs, C., Baba, J., 2014, PASA, 31, 35 \\

Dominguez Sanchez, H., Huertas-Company, M.,
Bernardi, M., Tuccillo, D., Fischer, J. L., 2018, MNRAS, 476, 3661 \\

D'Onghia, E., Vogelsberger, M., Hernquist, L., 2013, ApJ, 766, 34 \\


Egusa, F.,  Mentuch Cooper,  E., Koda, J., Baba, J.,
2017, MNRAS, 465, 460 \\

Elmegreen, D. M., 1981, ApJS, 47, 229 \\

Elmegreen, D. M., Elmegreen, B. G., 1984, ApJS, 54, 127 \\

Fujii, M., et al. 2011, ApJ, 730, 109 \\

Fujimoto, M., 1968, ApJ, 152, 391  \\

Garcia-Gomez, C., et al., 2017, A\&A, 601, 132 \\

Grand, R. J. J., et al. 2013, A\&A, 553, 77 \\

Hart. R. E., et al., 2017, MNRAS, 472, 2263 \\

Hewitt, I. B., Treuthardt, P., 2020, MNRAS, 493, 3854 \\

Huertas-Company, M., et al. 2015, ApJS, 221, 8 \\

Iye, M., et al. 2019, ApJ, 886, 1331 \\

Kendall, S., et al., 2008, MNRAS, 387, 1007 \\

Kennicutt, R. C. Jr. 1981, AJ, 86, 1847 \\

Koribalski et al., 2020, Ap\&SS, 365, 118 \\

Lintott, C. J., et al. 2008, MNRAS, 389, 1179 \\

Ma, J., 2001, ChJAA, 1, 395 \\

Mastes, K. L., et al., 2019, MNRAS, 487, 1808 \\

Pettitt, A. R., et al. 2020, MNNRAS, 498, 1159 \\

Rix, H.-W.,  Zaritsky, D., 1995, ApJ, 447, 82 \\

Roberts, W. W., 1969, AJ, 158, 123 \\

Ronneberger, O., Fischer, P., Brox, T., 2015,
Medical Image Computing and Computer-Assisted 
Intervention (MICCAI), Springer, LNCS, Vol.9351: 234 \\

Seigar, M. S., James, P. A., 1998, MNRAS, 299, 685 \\

Seigar, M. S., et al., 2008, MNRAS, 389, 1911 \\

Sellwood, J. A., 2011, MNRAS, 410, 1637 \\

Sellwood, J. A., Carlberg, R. G., 1984, ApJ, 282, 61 \\

Sellwood, J. A., Binney, J. J., 2002, MNRAS, 336, 785 \\

Shah, M., et al., 2019, MNRAS, 482, 4188 \\

Shen, A. X., Bekki, K., 2020, MNRAS, 497, 5090 \\

Tadaki, K., et al. 2020, ApJ, 901, 74 \\

Tasker E. J., Wadsley J., Pudritz R., 2015, ApJ, 801, 33 \\

\end{thebibliography}
%

\end{document}